\documentclass[12pt]{iopart}

\usepackage{graphicx}
\usepackage{iopams}
\usepackage{matutil}
\usepackage{subfigure}
\usepackage{harvard}

\begin{document}

\title{Logarithmically slow onset of synchronization}
\author{Gil Benk{\"o} and Henrik Jeldtoft Jensen}
\address{Institute for Mathematical Sciences, Imperial College London,\\
53 Prince's Gate, South Kensington Campus, SW7 2PG, London, UK\\
and}
\address{
Department of Mathematics, Imperial College London,\\
South Kensington Campus, SW7 2AZ, London, UK}
\eads{\mailto{g.benkoe@imperial.ac.uk}, \mailto{h.jensen@imperial.ac.uk}}


\begin{abstract}
Here we investigate specifically the transient of a synchronizing system, considering synchronization as a relaxation phenomenon. 
The stepwise establishment of synchronization is studied in the system of dynamically coupled maps
introduced by Ito~\&~Kaneko (2001 \textit{Phys.\ Rev.\ Lett.} \textbf{88} 028701, 2003 \textit{Phys.\ Rev.~E} \textbf{67} 046226), where the plasticity of dynamical couplings might be relevant in the context of neuroscience.
We show the occurrence of logarithmically slow dynamics in the transient of a fully deterministic dynamical system.
\end{abstract}


\section{Introduction}
The importance of transients in the dynamics of complex systems is manifold~: for example,
the transient state can be more relevant than the equilibrium state if most or more of the time is spent in the former. Also, relaxation dynamics can inform on the underlying energy, fitness, or cost landscape of a system \cite{Wales2004}, and thus help to understand it better as a whole.

In this paper we are interested in studying synchronization, a dynamical property of networks that is widely observed in fields such as 
optics, chemistry, biology and ecology, for instance in the brain \cite{Gray1989,Varela2001} 
 or in fireflies. Synchronization has been analysed for many physical
systems \cite{Pecora1997,Pikovsky2001}. 
The onset of synchronization is of general relevance. Similarity with relaxation, for example in spin glasses, or in superconductors, can be used to study synchronization. Transients can be used to probe the properties of synchronizing systems \cite{Arenas2006}.
However, there has been little research so far on synchronization as a relaxation phenomenon \cite{Abramson2000,Manrubia2000}.
 The aim of this paper is to systematically study the transient of a synchronizing system.

A specific form of relaxation dynamics are glassy dynamics, when the transient is extremely long \cite{Jensen2007}.
Furthermore, in a number of systems with glassy dynamics the special case of logarithmically slow dynamics has been observed.
However, so far all of these systems were stochastically driven.
Here we show the occurrence of logarithmically slow dynamics in the transient of a fully deterministic dynamical system.
In the first part of the paper we explain the model, a globally coupled map (GCM), with adaptive coupling which is inspired by the plasticity of synapses. We then study characteristics of its transients for a range of parameters.
Finally we show that for some parameters the transient is logarithmically slow and that it can be explained by a simple model.

\section{The Ito-Kaneko model of synchronization}

The Ito-Kaneko model \cite{Ito2001,Ito2003} is a globally coupled map (GCM), a coupled simultaneous system of $N$ logistic equations. The individual maps or units $x^i$ are defined and coupled as follows :

\begin{eqnarray}\eqalign{
x^i_{t+1} & =  \left(1-c\right) f(x^i_t) + c \sum^N_{j=1} w^{ij}_t f(x^j_t) \label{eq:gcm1} \\
f(x) & =  a x \left(1-x\right) \\
w^{ij}_{t+1} & =  \frac{[ 1 + \delta g(x^i_t,x^j_t) ] w^{ij}_t}
        {\sum^N_{j=1} [ 1 + \delta g(x^i_t,x^j_t) ] w^{ij}_t} \\
g(x,y) & =  1 - 2 \left|x-y\right|  \,,}
\end{eqnarray}

\noindent where $a$ is the logistic equation parameter and $c$ is the coupling
parameter. The coupling is further tuned by weights $w^{ij}$, which are dynamical variables as well.
The function $g$, scaled by a parameter $\delta$, defines a Hebbian update of the connection weights, by reinforcing the connections between similar
units. This Hebbian plasticity of the couplings is inspired by
the synaptic plasticity which enables nervous systems to learn \cite{Antonov2003}.

\begin{figure}
  \subfigure{\includegraphics[width=.99\textwidth]{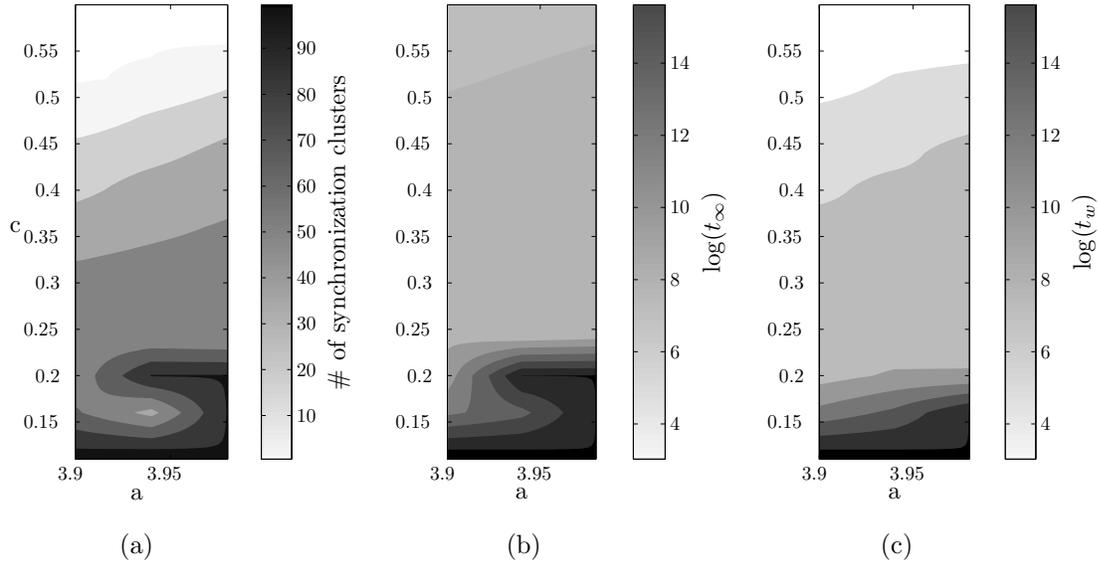}} \\[-3ex]
  \setcounter{subfigure}{0}
  \hspace*{.095\textwidth}
	\subfigure[]{\label{fig:p1}}
	\hspace{.31\textwidth}
	\subfigure[]{\label{fig:p2}}
	\hspace{.3\textwidth}
	\subfigure[]{\label{fig:p3}}
  \caption{(a) Part of the phase diagram of the Ito-Kaneko model (b) Length of transient $t_\infty$ 
(c) Time $t_w$ until the variation of weights falls below 0.1\% (all averaged over 25 realizations).
}\label{fig:pdpart}
\end{figure}

The system exhibits three different long-term behaviours which can
be classified into three phases, depending on the parameters. In
the coherent phase $C$, all units synchronize, forming one
synchronized cluster containing all units. In the ordered
phase $O$, the set of units is partitioned into subsets or clusters
$C_k$ within which there is synchronization or which contain
single units not synchronized with any other unit. In the following, in line with
 \cite{Ito2001,Ito2003,Manrubia2000}, we will call the number of parts
in this partition the cluster number. Finally, in
the disordered phase $D$, no synchronization at all is achieved,
forming a partition with a cluster number of  value $N$. The phase
diagram in \fref{fig:p1} shows the boundaries between
predominant phases in a part of the parameter space.

\section{Synchronization transients}
%
%
It has been found previously that in a static links version of the Ito-Kaneko model its transient is exponential in the ordered regime $O$ and that its length diverges when the border $O/D$ between the ordered and disordered regimes is  approached \cite{Abramson2000,Manrubia2000}.
Stretched exponential decay of correlation functions has been found in a similar system \cite{Katzav2005}.

In the following we study the transient in detail, and focus on the way events during the transient are simulated and detected.
The system \eref{eq:gcm1} was simulated numerically, each initial $x^i$ being randomly chosen in the interval $[0,1]$.
The initial $w^{ij}$ are set to $1/(N-1)$.
A higher precision simulation method similar to Pikovsky \emph{et~al.}~\cite{Pikovsky2001a} was used
in order to avoid synchronization artifacts due to limited numerical precision.
Two units $x^i$ and $x^j$ were considered to be synchronized if $|x^i - x^j| < 10^{-270}$.
Due to the resulting increased computation times we focused on a part of the parameter space, shown in \fref{fig:pdpart}. 
During each run of a simulation, the time steps $t_k$ at which two units synchronized were recorded. We term these $t_k$ synchronization events. 
As the successive synchronization is analyzed in analogy to the investigation of avalanches and quakes in  \cite{Anderson2004}, where a set of events is considered as one quake, here any synchronization events less than $\Delta t=10$ time steps apart are considered together as a single synchronization event.

In \fref{fig:nbvst}(a) we show as an example the time evolution of the number of synchronized clusters for a single simulation run. The number of clusters drops stepwise from $N = 100$ during a long transient.
We note $N_{\textrm{sync}}(t)$ the number of synchronization events until time step $t$. If, after a synchronization event at time step $t_k$, no further synchronization events happened for a duration of $4 t_k$, the simulation was stopped. We define the synchronization transient as the time from the beginning of the simulation until the last synchronization event.

\begin{figure}
  \centering
  \includegraphics[width=.9\textwidth]{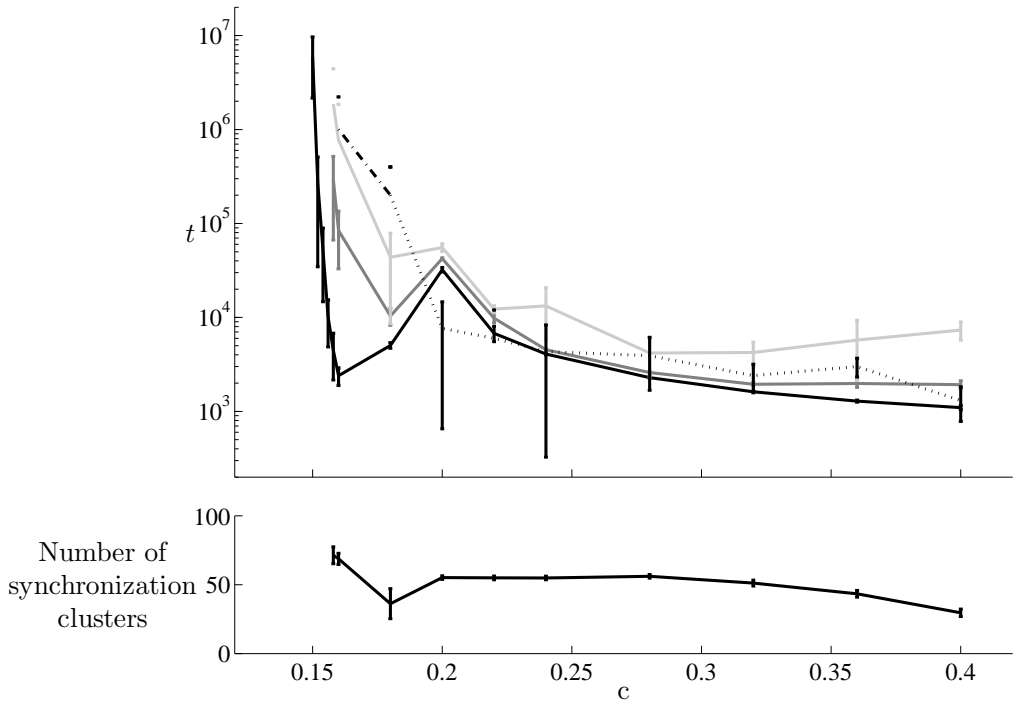}\\
  \caption{(Top) Divergence of transient length as the border to the disordered region is approached, with $a = 3.9$\,. Shown are the time of the first synchronization event $t_1$ (solid black), of the 25th synchronization event $t_{25}$ (dark gray), and of the last synchronization event $t_\infty$ 
(light gray), and $t_w$ (dotted black when $w$ converges in all realizations, dash-dotted otherwise). (Bottom) Corresponding number of synchronization clusters (averaged over 25 simulation runs).}\label{fig:firstrec}
\end{figure}

The dependence of the transient on the parameters $a$ and $c$ is shown in \fref{fig:pdpart}(b) and~(c). The main feature is that the length of the transient, and also the time until the connection weights $w^{ij}$ stabilize, increases as the the overall coupling $c$ decreases.
\Fref{fig:firstrec} shows again in detail how the length of transient diverges as $c$ diminishes, with the parameter of the logistic function $a = 3.9$ constant. This makes intuitively sense as the system changes from a regime with some synchronization (ordered) to a regime with no synchronization at all (disordered). For the rest of this paper, $a$ is fixed to 3.9\,.

A theoretical value for the coupling $c_{O/D}$ at which the length of the transient diverges can be analytically derived for a system in which the weights $w^{ij}$ stabilize, see  \cite{Ito2001,Ito2003}.
The divergence of the transient corresponds to the border between the ordered and disordered regimes and can thus be calculated by studying the stability of the ordered state, using the transversal Lyapunov exponent. At the considered border the system only forms synchronization clusters of size up to 2, i.e.\ only isolated pairs of synchronized units are formed. The transversal Lyapunov exponent for a system in which the weights stabilize is then \cite{Ito2001,Ito2003}:
\begin{equation}\label{eq:transexp}
\Lambda_\bot = \ln|1 - 2c| + \Lambda_f \,,
\end{equation}
where $\Lambda_f$ is the Lyapunov exponent of $f(x)$, the logistic function. For $a = 3.9$\,, the Lyapunov exponent of the logistic function calculated by simulation is $\Lambda_f \approx 0.73$ and we obtain a theoretical value $c \approx 0.26$\,.
In our system the weights $w^{ij}$ do not stabilize, which seems to enable synchronization at lower $c$ values, as the border lies at $c_{O/D} \approx 0.15$\,.

\begin{figure}
  \subfigure{\includegraphics[width=.99\textwidth]{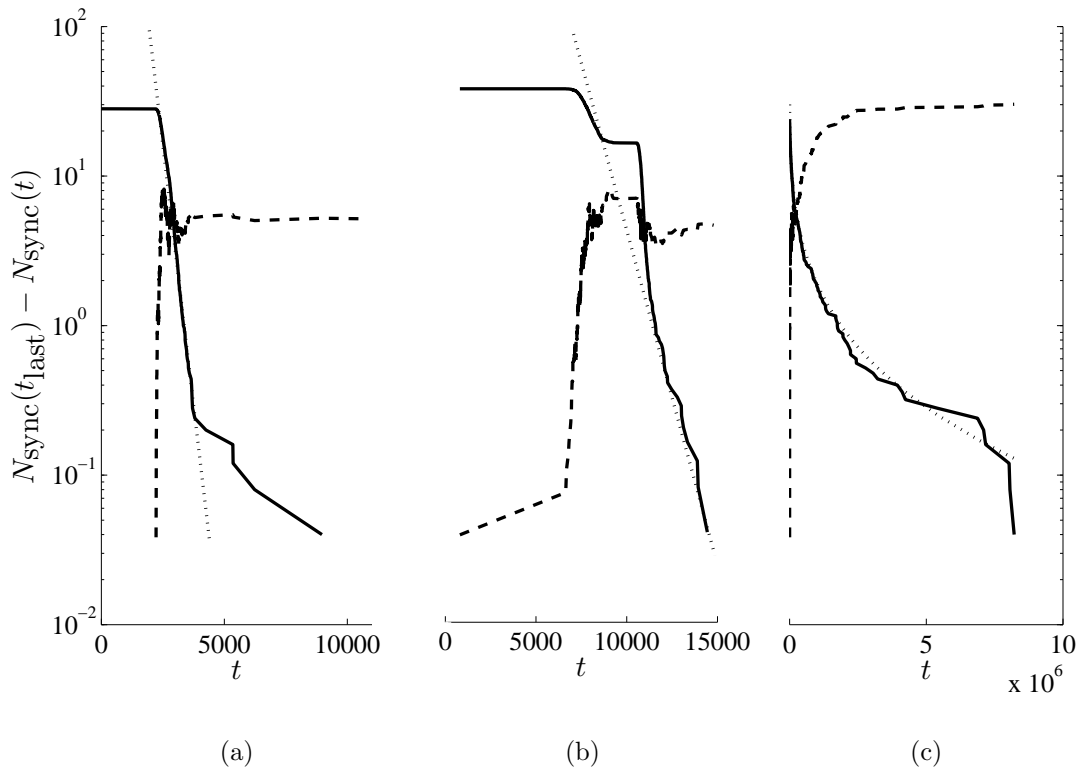}} \\[-3ex]
  \setcounter{subfigure}{0}
  \hspace*{.175\textwidth}
	\subfigure[]{\label{fig:n1}}
	\hspace{.275\textwidth}
	\subfigure[]{\label{fig:n2}}
	\hspace{.275\textwidth}
	\subfigure[]{\label{fig:n3}}
  \caption{Time dependence of $N_{\textrm{sync}}(t_\infty)-N_{\textrm{sync}}(t)$ averaged over 25 simulation runs (solid) with variance (dashed) and fitted exponential or stretched exponential of the form $\exp({-t^{0.28}})$ (dotted), $c =$ 0.28, 0.22, 0.158 (from left to right).}\label{fig:nb3vst} 
\end{figure}

We studied the transient in a logarithmic timescale for $c$ values close to and far from $c_{O/D}$.
In \fref{fig:nb3vst} we plotted $N_{\textrm{sync}}(t_\infty)-N_{\textrm{sync}}(t)$ against time in order to detect exponential transients of the form $N_{\textrm{sync}}(t_\infty)-N_{\textrm{sync}}(t) \propto \rme^{-\beta t}$, as described in  \cite{Abramson2000,Manrubia2000}.
For $c = 0.28$\,, shown in \fref{fig:n1}, the transient is indeed exponential. However, as $c$ decreases, the transient turns more irregular, see \fref{fig:n2} ($c = 0.22$). Eventually, for $c = 0.158$\,, very close to $c_{O/D}$, the transient is extremely long and settles into the stretched exponential shape shown in \fref{fig:n3}.
Stretched exponential relaxation has been observed in related systems \cite{Katzav2005}.
Also the evolution of the variance of the considered quantity is different depending on whether $c$ away and very close to the border.
The difference in the shape of the transient and its variance indicates that a different process is underlying the system at the border $O/D$. In the following we study further the transient statistics at the border $O/D$.

\section{Logarithmically slow transients}

We are especially interested in the unusual dynamics at the border $O/D$.
Also, this border separates ordered from disordered behaviour and is interesting because of the relevance of computation at the edge of chaos in neural nets \cite{Natschlaeger2005}.
The transient at the border is characterized by its extreme length.

Extremely long transients have also been observed in other many component systems with glassy dynamics \cite{Jensen2007}.
There, the time span needed to reach a steady, time independent state is often far beyond experimentally accessible time scales. For example, when melted alloys are cooled down they typically retain the amorphous arrangement characteristic of the liquid high temperature phase while the molecular mobility decreases many orders of magnitude, rendering it near impossible to reach thermodynamic equilibrium. However, over short time scales the system properties may appear to be time independent as in thermal equilibrium. Only when several orders of magnitude of time scales are covered the slow change of macroscopic characteristic properties with time can be resolved directly.

\begin{figure}
  \centering
  \includegraphics[width=.9\textwidth]{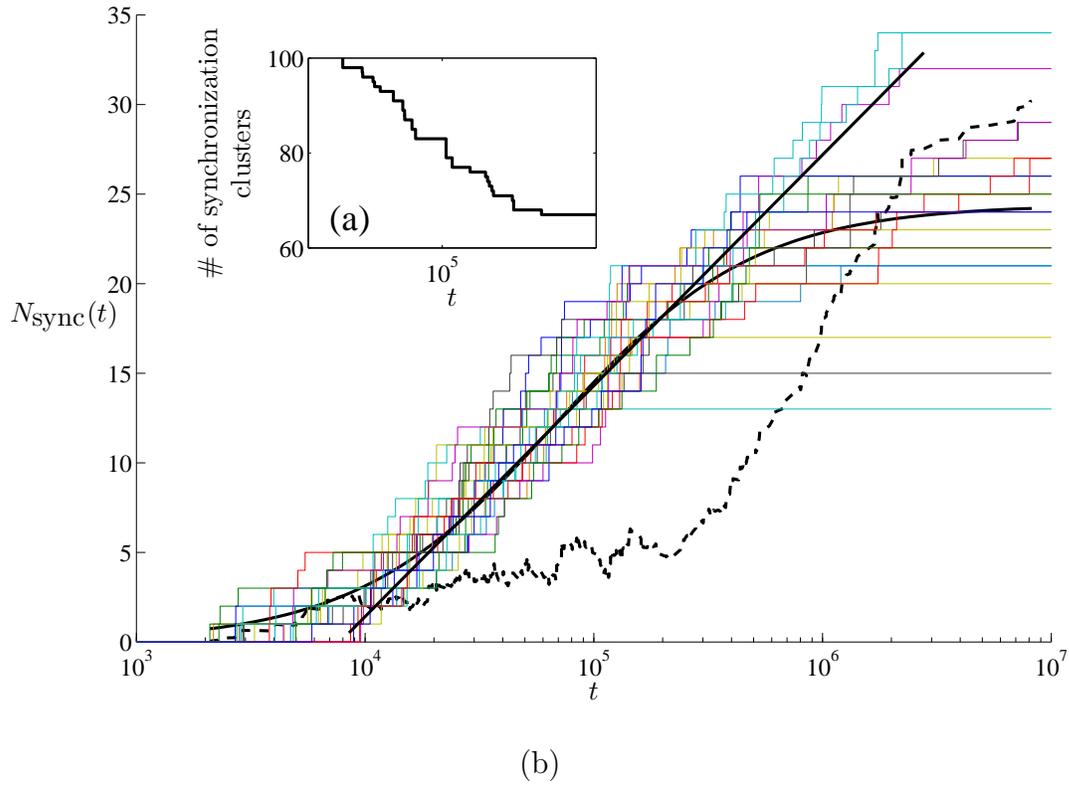}\\
  (b)\\
  \caption{Synchronization transient near the border $O/D$, with $c = 0.158$\,. (a-insert) Time dependence of the number of synchronization clusters for a single simulation run. (b) Time dependence of the number of synchronization events for 25 simulation runs (thin lines) and corresponding variance (dashed), a linear fit and a fit of the simple model described by \eref{eq:synctr} (thick lines).}\label{fig:nbvst}
\end{figure}

A superposition of 25 simulations runs for $c = 0.158$ close to $c_{O/D}$ covering the whole very long transient up to $t = 10^7$ is shown in \fref{fig:nbvst}. Due to computational complexity no value closer to $c_{O/D}$ was used. Interestingly, the transient is linear on a logarithmic timescale for intermediate times. In the following we describe an approach to modeling the transient.

A simple model would be to assume that the rate of synchronization events is proportional to the rate at which any two units have close values by chance, i.e.~'collide'. However, at low overall coupling $c$, units $x^i$ can only synchronize with a single additional unit and form pairs. Thus units that are already paired up are not available anymore to synchronize with other units. So simplifying further, if we note the number of synchronization events $N_{\textrm{sync}}(t)$, and the number of unsynchronized units is approximately $N - N_{\textrm{sync}}(t)$, we obtain:
    \begin{equation}
    \frac{\rmd N_{\textrm{sync}}(t)}{\rmd t} = k (N - N_{\textrm{sync}}(t))^2
    \end{equation}
and
    \begin{equation}
    N_{\textrm{sync}}(t) = N - \frac{1}{\frac{1}{N}+k t}\,.
    \label{eq:synctr}
    \end{equation}
We further assume that only an effectual subset of the $N$ units can actually synchronize due to the low overall coupling $c$, so we replace $N$ by an effectual, i.e.~actually operating number of units $N_\textrm{eff}$ in the above formula.
The predicted $N_{\textrm{sync}}(t)$ agrees well with the simulated data for $N_\textrm{eff} = 24.5$ and $k = 6\cdot10^{-7}$, see \fref{fig:nbvst}(b). However, it is not possible to find a reasonable fit of the simulated $N_{\textrm{sync}}(t)$ to \eref{eq:synctr} for higher values of $c$.

We also observed that $N_{\textrm{sync}}(t)$ is linear on a logarithmic timescale for intermediate times. This might be related to the log-time dependence that has been previously obtained for non-homogeneous Poisson processes in logarithmic time (log-Poisson). The hypothesized mechanism behind these processes is inspired by intermittency studies of fluctuations in glassy systems, that have demonstrated that large intermittent fluctuations are responsible for the deviations from equilibrium statistics \cite{Buisson2003}. It was suggested that abrupt and irreversible moves from one metastable configuration to another, so-called quakes, are a result of record sized fluctuations. The assumption that the metastable attractors typically selected by the glassy dynamics have marginally increasing stability  means that a fluctuation bigger than any previously occurred fluctuation, i.e. a record-sized fluctuation, can induce a quake \cite{Sibani2003,Sibani1993}. Quakes lead to entrenchment into gradually more stable configurations, and carry the average drift of the dynamics. The quakes have a similar effect on a logarithmic time scale, which might be modeled by a Poisson process in logarithmic time.

Log-Poisson processes have been observed in the NK model of evolution \cite{Sibani99}, 
charge-density waves \cite{Sibani1993}, and further in spin glasses, supercooled magnet relaxation, and the Tangled Nature evolution model \cite{Anderson2004}.
While in a Poisson process the probabilities for events $t_k$ are characterized by \cite{vanKampen2007}:
\begin{equation}
P[ N(t+\tau) - N(t) = k ] =
\frac{\rme^{-\lambda\tau}(\lambda\tau)^k}{k!}
\end{equation}
and
\begin{equation}
P[ t_k - t_{k-1} < x ] = 1 - \rme^{-\lambda x}\,,
\end{equation}
in analogy, in a log-Poisson process we have \cite{Sibani2003}:
\begin{equation}
P[ N(t+\tau) - N(t) = k ] =
\frac{1}{k!}\left(\frac{t+\tau}{t}\right)^{-\lambda}\left(\lambda\frac{t+\tau}{t}\right)^k
\end{equation}
\begin{equation}
P[ \ln( t_k/t_{k-1} ) < x ] = 1 - \rme^{-\lambda x} \,.
\label{eq:cdtt}
\end{equation}
In both cases, it is easy to show that the PDF of the number of events $N(t)$ and its variance are equal.
Thus in \fref{fig:nb3vst} $N_{\textrm{sync}}(t_\infty)-N_{\textrm{sync}}(t)$ and its variance  should be symmetric, which is observed indeed only up to some error on the border $O/D$ (\fref{fig:n3}) while away from the border (\fref{fig:nb3vst}(a) and (b)) $N_{\textrm{sync}}(t_\infty)-N_{\textrm{sync}}(t)$ and its variance seem unrelated.

\begin{figure}
  \centering
  \subfigure{\label{fig:t1}}
  \subfigure[]{\includegraphics[width=.85\textwidth]{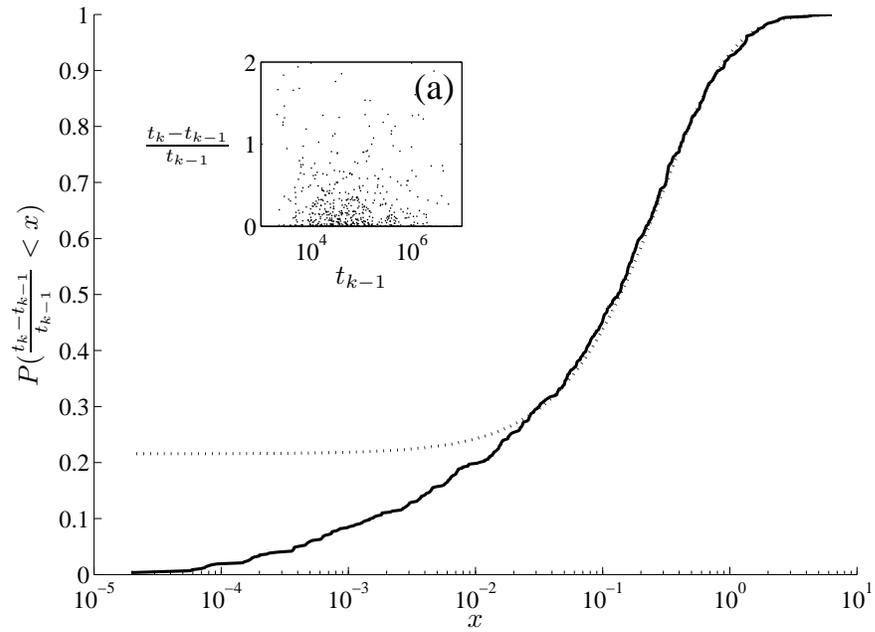}\label{fig:t2}}\\
  \subfigure[]{\includegraphics[width=.85\textwidth]{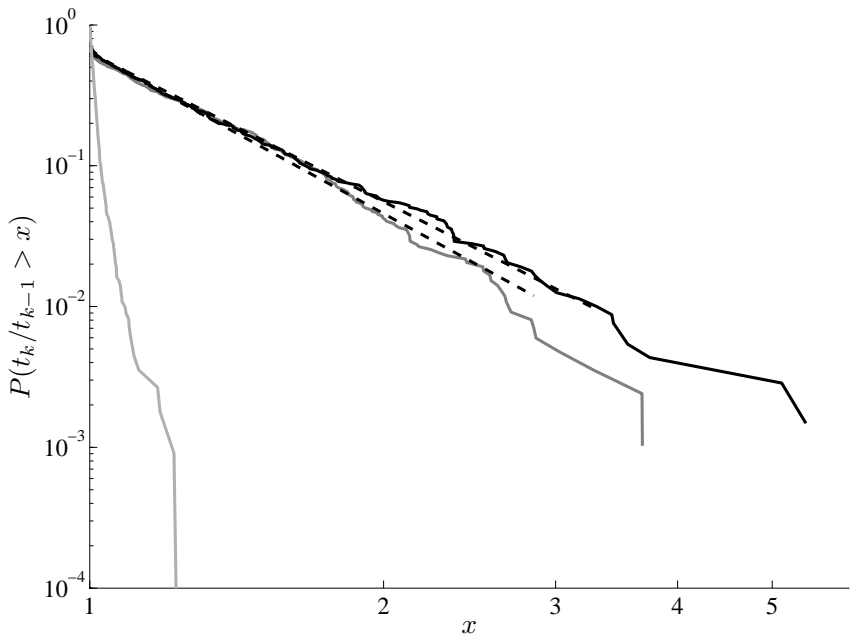}\label{fig:t3}}
  \caption{ (a-insert) Distribution, (b) cumulative distribution of $({t_k-t_{k-1}})/{t_{k-1}}$ for $c= 0.158$ (solid), and theoretic distribution for a log-Poisson process (dotted) (c) Cumulative distribution of ${t_k}/{t_{k-1}}$ close to the border $O/D$ with $c=0.2$ (light gray), 0.16 (dark gray) and 0.158 (black), and fitted theoretic distribution for a log-Poisson process (dashed). }\label{fig:ttt}
\end{figure}

Thus the process described by \eref{eq:synctr} seems to be an alternative way to obtain log-time dependence in an intermediate time regime.
We can further corroborate this finding by studying the cumulative distribution of $({t_k-t_{k-1}})/{t_{k-1}}$, see \fref{fig:t2}. For a static Poisson process, this distribution would be a step function. Instead, it agrees well with the theoretic distribution for a log-Poisson process, $P(({t_k-t_{k-1}})/{t_{k-1}} < x) = 1-(1+x)^{-\lambda}$ with $\lambda = 3.5$\,. The support for the distribution at low $({t_k-t_{k-1}})/{t_{k-1}}$ comes from the initial phase of the transient, where the initial conditions dominate, 
which might explain the discrepancy between the theoretical and simulated distribution there.
Also, as shown in \fref{fig:t1}, the value of $({t_k-t_{k-1}})/{t_{k-1}}$ stays within a relatively small range over a range of orders of magnitude of $t_k$, while it would quickly drop to zero in a static Poisson process.
In \fref{fig:t3} we show the cumulative distribution of $t_k/t_{k-1}$. This distribution also agrees well with the theoretical cumulative distribution \eref{eq:cdtt} for a log-Poisson process with $\lambda = 3.5$ and seems to corroborate that the synchronization process in this model is an alternative way to obtain logarithmically slow relaxation dynamics.

\begin{figure}
  \centering
  \subfigure[]{\includegraphics[width=.85\textwidth]{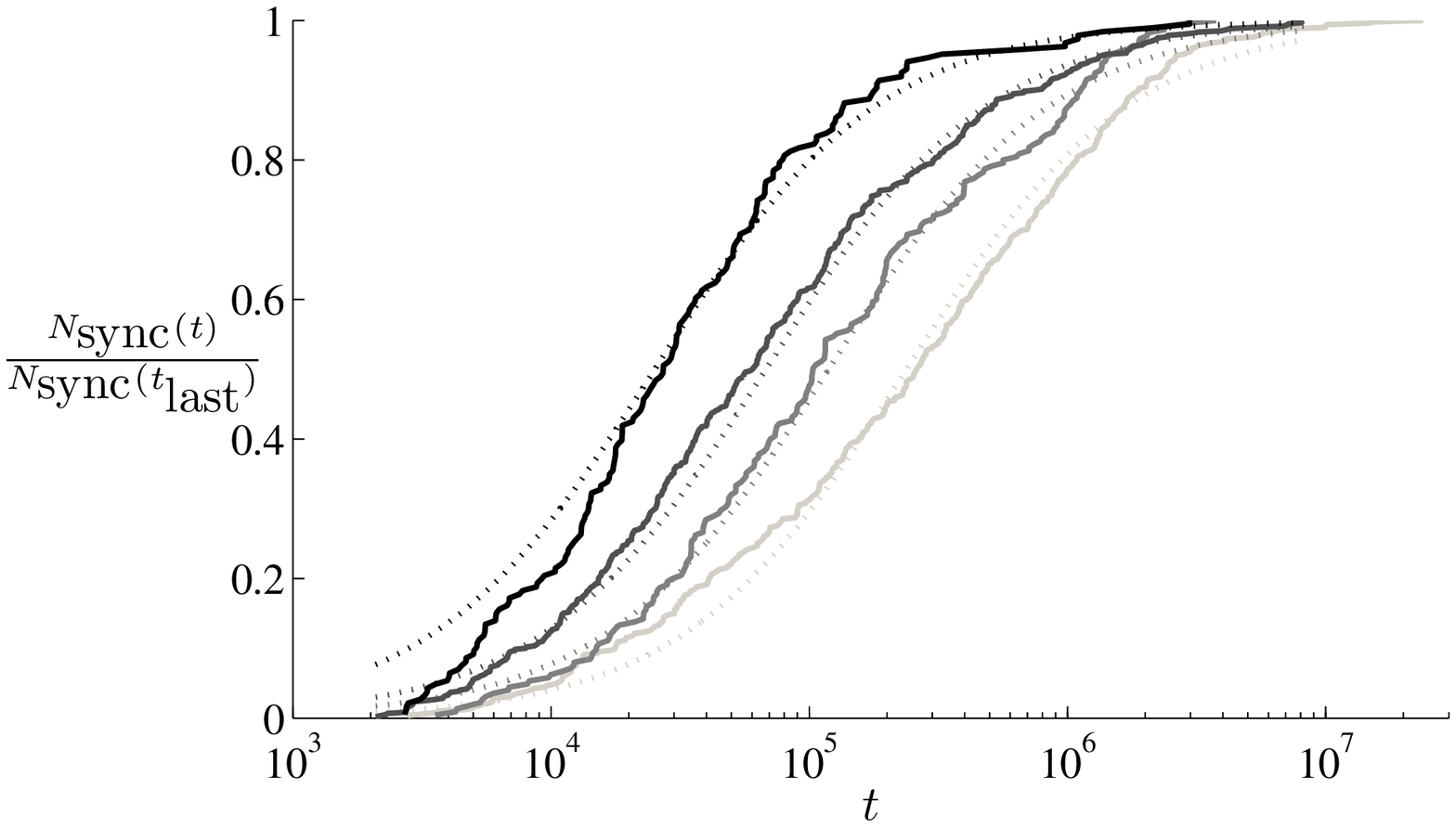}\label{fig:s1}}\\
  \subfigure[]{\includegraphics[width=.85\textwidth]{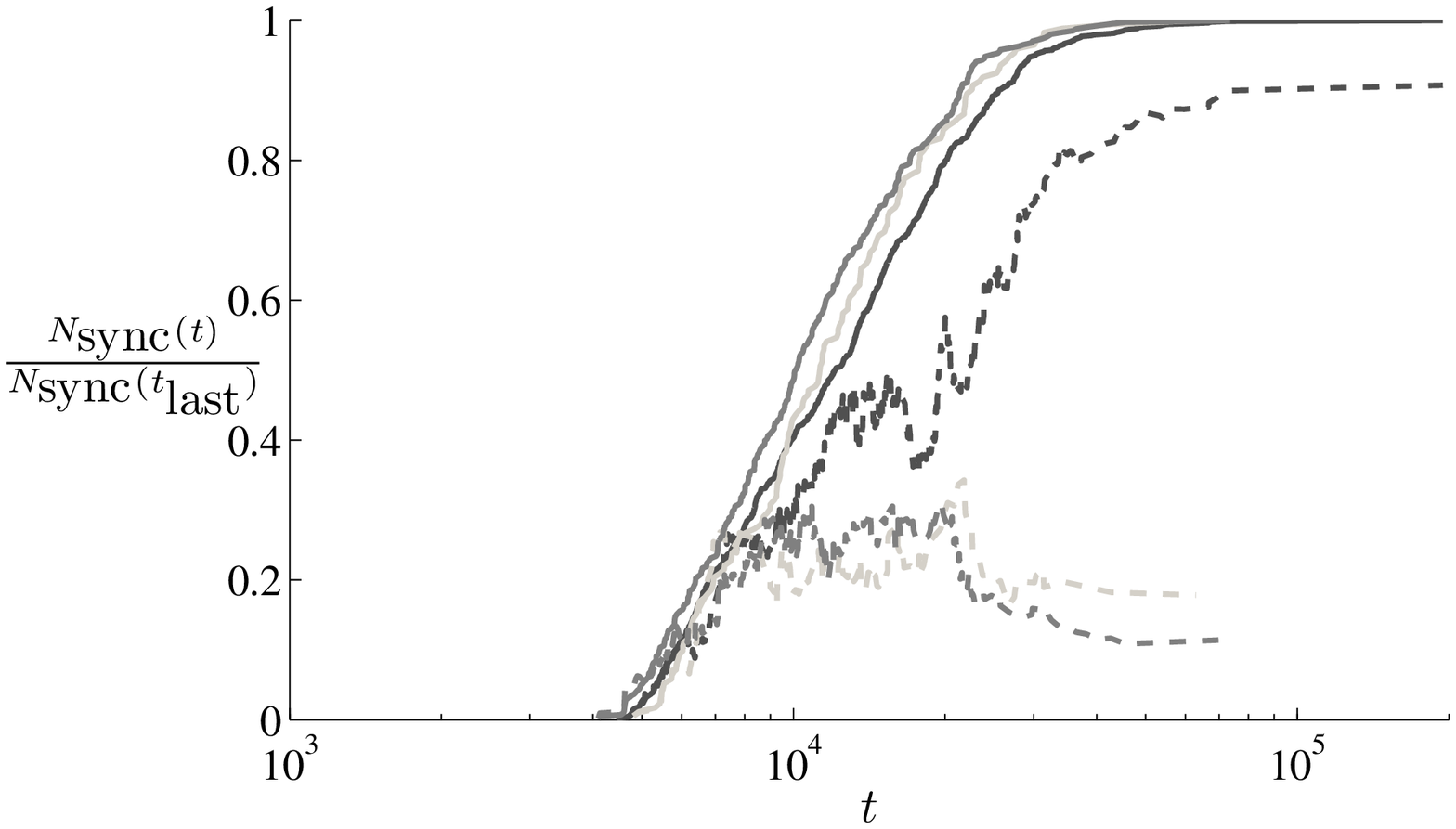}\label{fig:s2}}
  \caption{ System size dependence of the transient. (a) Transients for $c=0.158$ and $N=25,50,100,$ and 200 (solid gray to black) and corresponding theoretical transients using \eref{eq:synctr}, $N_{\textrm{eff}}=7,14,24.5,$ and 61 and  $k = 6\cdot10^{-7}$ (dotted). (b) Transients for $c=0.18$ and $N=25,50,$ and 100 (solid) and variance (dashed). }\label{fig:size}
\end{figure}

System size can be relevant to the shape of the transient of a system. A superposition of the transients for different system sizes, normalized by the final 
number of synchronization events, and for $c=0.158$ close to $c_{O/D}$ and $c=0.18$, is shown in \fref{fig:size}(a) and~(b).
Due to computational complexity, the sampling size is small, especially for $N=200$\,.
The simulations show that the system size dependence close to $c_{O/D}$ seems to be a shifting of the transient to faster synchronization. The shifting is close to what \eref{eq:synctr} suggests.
Also, for $c=0.18$\,, further away from $c_{O/D}$, the transients are almost equivalent for different system sizes when normalized by the final 
number of synchronization events. Thus also here the system synchronizes faster for bigger system sizes.
The faster synchronization for increasing system size can be rationalized using the same idea that led to \eref{eq:synctr}, as follows: since the values of the units $x^i$ are constrained to the interval $[0,1]$ 
of the logistic map, increasing the system size increases the number of units in $[0,1]$ and therefore increases the number of collisions, i.e.~the speed of synchronization.

\section{Conclusion}

In conclusion, we have studied synchronization as a relaxation phenomenon. The transient of synchronization in a coupled map model was found to drastically depend on the amount of overall coupling. For an overall coupling at the border to disorder, the transient was found to be logarithmically slow at intermediate times. This behaviour has been found in other systems exhibiting record statistics, where the dynamics is termed log-Poisson and arises because record fluctuations become increasingly rare. In the model of the present system the dynamics may arise by an alternative way : it can be simply explained by (for intermediate times) it getting more and more difficult to find a new synchronization partner and actually synchronize as the system ages. Interestingly, this kind of dynamics has been found otherwise in noise-driven systems. The coupled map model used in this paper has no noise term, however, it uses the logistic map, which has been studied as a noise generator since Ulam \& Neumann \cite{Ulam1947}. Thus the use of the logistic map might be an ingredient for the appearance of logarithmically slow dynamics. In a similar system it was shown that a coupled map lattice is equivalent to a system of stochastic PDE \cite{Katzav2005}.

One example for the relevance of the synchronization observed in GCM is neural networks.
The speed of synchronization in neural networks is important \cite{Kopell2004}, 
as it presumably is related to the response time of the brain. Namely, if we assume, along with Varela \emph{et~al.}~\cite{Varela2001}, that the state of the brain is given by its state of synchronization, then the time to swap from one synchronization pattern to another seems to determine how fast the brain can react.
Also, the
behaviour at the edge between partial synchronization and disorder is especially interesting because of the relevance of computation at the edge of chaos \cite{Natschlaeger2005}.

\ack{}
Fruitful comments by Paolo Sibani, discussions with Adele Peel, computer support by Andy Thomas and the use of the Imperial College High Performance Computing Service 
are gratefully acknowledged.

\section*{References}

\bibliographystyle{jphysicsB}
\bibliography{recordk}

\end{document}